\documentclass{emulateapj}

\usepackage{natbib}
\usepackage{amsmath}

\newcommand{\ugriz}{\protect\hbox{$ugriz$} }

\newcommand{\about}{$\sim\!\!$~}

\newcommand{\err}[2]{\ensuremath{^{+#1}_{-#2}}}

\def\lsim{\hbox{\rlap{\raise 0.425ex\hbox{$<$}}\lower 0.65ex\hbox{$\sim$}}}
\def\gsim{\hbox{\rlap{\raise 0.425ex\hbox{$>$}}\lower 0.65ex\hbox{$\sim$}}}

\shorttitle{Galsnid}
\shortauthors{Foley \& Mandel}

\begin{document}

 \title{Classifying Supernovae Using Only Galaxy Data}

\def\illast{1}
\def\illphy{2}
\def\cfa{3}

\author{
{Ryan~J.~Foley}\altaffilmark{\illast,\illphy} and
{Kaisey~Mandel}\altaffilmark{\cfa}
}

\altaffiltext{\illast}{
Astronomy Department,
University of Illinois at Urbana-Champaign,
1002 W.\ Green Street,
Urbana, IL 61801 USA
}
\altaffiltext{\illphy}{
Department of Physics,
University of Illinois Urbana-Champaign,
1110 W.\ Green Street,
Urbana, IL 61801 USA
}
\altaffiltext{\cfa}{
Harvard-Smithsonian Center for Astrophysics,
60 Garden Street, 
Cambridge, MA 02138 USA
}

\begin{abstract}
  We present a new method for probabilistically classifying supernovae
  (SNe) without using SN spectral or photometric data.  Unlike all
  previous studies to classify SNe without spectra, this technique
  does not use any SN photometry.  Instead, the method relies on
  host-galaxy data.  We build upon the well-known correlations between
  SN classes and host-galaxy properties, specifically that
  core-collapse SNe rarely occur in red, luminous, or early-type
  galaxies.  Using the nearly spectroscopically complete Lick
  Observatory Supernova Search sample of SNe, we determine SN
  fractions as a function of host-galaxy properties.  Using these data
  as inputs, we construct a Bayesian method for determining the
  probability that a SN is of a particular class.  This method
  improves a common classification figure of merit by a factor of
  $>$2, comparable to the best light-curve classification techniques.
  Of the galaxy properties examined, morphology provides the most
  discriminating information.  We further validate this method using
  SN samples from the Sloan Digital Sky Survey and the Palomar
  Transient Factory.  We demonstrate that this method has wide-ranging
  applications, including separating different subclasses of SNe and
  determining the probability that a SN is of a particular class
  before photometry or even spectra can.  Since this method uses
  completely independent data from light-curve techniques, there is
  potential to further improve the overall purity and completeness of
  SN samples and to test systematic biases of the light-curve
  techniques.  Further enhancements to the host-galaxy method,
  including additional host-galaxy properties, combination with
  light-curve methods, and hybrid methods should further improve the
  quality of SN samples from past, current, and future transient
  surveys.

\end{abstract}



\section{Introduction}\label{s:intro}

Supernovae (SNe) are the result of several different kinds of
explosions from different stellar progenitor systems.  Separating SNe
into different classes (or ``types'') has existed for over 70 years
\citep{Minkowski41}.  Despite the recent discovery of several new
varieties of SNe \citep[e.g.,][]{Foley13:iax}, most SNe discovered can
be placed into the two broad categories of ``core-collapse'' SNe
(those with a massive star progenitor, corresponding to the Ib, Ic,
II, IIb, and IIn classes) and Type Ia SNe (SNe~Ia).
\citet{Filippenko97} reviews the observational characteristics of
these more common classes.  There remain a few additional SNe that do
not fall into these two categories, but these are a small fraction of
the SNe discovered \citep{Li11:rate2}.

Current transient surveys discover many more SNe than can be
spectroscopically classified.  Future surveys will have even lower
rates of spectroscopic classification.  Because of this limitation,
there has been a significant amount of effort to photometrically
classify SNe \citep[see][for a review]{Kessler10}.  Providing large
and pure samples of individual classes of SNe is important for many
studies.  In particular, large samples of SNe~Ia are required to make
progress in determining the nature of dark energy
\citep[e.g.,][]{Campbell13}, driving cosmic acceleration, which was
originally discovered through measurements of (mostly)
spectroscopically confirmed SNe~Ia \citep{Riess98:Lambda,
  Perlmutter99}.  Additionally, having some preliminary classification
to aid in spectroscopic follow-up can be useful for studies of all
classes of SNe.

Until now all effort has focused on classification using only the
light curves of the SNe.  Specifically, different SN classes tend to
have different rise times, decline rates, and colors.  Additionally,
all efforts have focused on separating SNe~Ia from all other types of
SNe.  This problem becomes more difficult with low signal-to-noise
ratio data, sparse sampling, a sample that extends over a large
redshift range, and limited filters.  Nonetheless, the best
photometric classification methods for a simulated SN sample is 96\%
pure while recovering 79\% of all SNe~Ia \citep{Kessler10, Sako11,
  Campbell13}.

Our approach here is to classify SNe without using any SN photometry.
This method uses the known correlations between host-galaxy properties
and SN classes.  Since core-collapse SNe (SNe~Ibc and II) have massive
star progenitors and SNe~Ia have WD progenitors, several host-galaxy
properties are correlated with SN type.  For decades, we have known
that core-collapse SNe explode almost exclusively in late-type
galaxies and are associated with spiral arms and \ion{H}{2} regions.
On the other hand, SNe~Ia explode in all types of galaxies and have no
preference for exploding near spiral arms.  These basic facts drive
the majority of our exploration.

Most of the host-galaxy data we use should be available for all
transient surveys.  We do not attempt to combine this classification
technique with photometric classifications (or attempt hybrid
approaches), and leave such implementation to future studies.

The manuscript is structured in the following way.
Section~\ref{s:fom} describes a figure of merit for determining the
quality of classification.  We introduce our SN samples in
Section~\ref{s:data} and discuss host-galaxy properties in
Section~\ref{s:galaxy}.  Our method is described in
Section~\ref{s:galsnid}, and we test the method in
Section~\ref{s:tests}.  We discuss our results, additional
applications, and future prospects in Section~\ref{s:disc} and
conclude in Section~\ref{s:conc}.

\section{Figure of Merit}\label{s:fom}

When evaluating non-spectroscopic identification techniques, one needs
a metric for comparison.  \citet{Kessler10} presents a figure of merit
(FoM) for such a comparison, with the focus on producing a large
sample of SNe~Ia with low contamination.  The FoM, $\mathcal{C}_{\rm
  Ia}$ is the product of the efficiency and pseudopurity of a sample
of SNe classified as SNe~Ia.  The efficiency is defined as
\begin{equation}
  \epsilon_{\rm Ia} = N^{\rm Sub}_{\rm Ia} / N^{\rm Tot}_{\rm Ia},
\end{equation}
where $N^{\rm Tot}_{\rm Ia}$ is the total true number of SNe~Ia in the
full sample and $N^{\rm Sub}_{\rm Ia}$ is the true number of SNe~Ia in
a subsample classified as SNe~Ia under some criterion.  The
pseudopurity is defined as
\begin{equation}
  PP_{\rm Ia} = \frac{N^{\rm Sub}_{\rm Ia}}{N^{\rm Sub}_{\rm Ia} +
    W^{\rm False}_{\rm Ia} N^{\rm Sub}_{\rm Non-Ia}},
\end{equation}
where $N^{\rm Sub}_{\rm Non-Ia}$ is the number of objects
misclassified as SNe~Ia in the selected subsample which are not truly
SNe~Ia, and $W^{\rm False}_{\rm Ia}$ is the weight given to adjust the
importance of purity on the FoM.  If $W^{\rm False}_{\rm Ia} = 1$, the
pseudopurity is simply the true purity, which is equivalent to the
probability of a SN in that subsample being a SN~Ia.
\citet{Campbell13} suggests that $W^{\rm False}_{\rm Ia} = 5$ is the
preferred value for creating a sample of SNe~Ia for the purpose of
measuring cosmological parameters; we will use that value throughout
this paper.

If one can perfectly reject non-transient sources from the subsample,
then $N^{\rm Sub}_{\rm Non-Ia} \approx N^{\rm Sub}_{\rm CC}$, where
$N^{\rm Sub}_{\rm CC}$ is the number of core-collapse SNe in the
subsample (there will perhaps be a few peculiar thermonuclear SNe in
the subsample, but those will likely be orders of magnitude smaller
than the core-collapse population).

Of course this FoM is not the only way to compare classification
methods, and it is particularly focused on selecting a relatively pure
sample of SNe~Ia.  But since we hope to merely provide practical and
useful methods of classification for various surveys, there is no
urgent need to define a different FoM.  Regardless of whether one is
trying to select SNe~Ia or core-collapse or wants to somehow weight
efficiency and/or purity differently, the \citet{Kessler10} FoM will
likely still be informative.

\section{Data}\label{s:data}

\subsection{Supernova Samples}

When testing various classification schemes, one would ideally have a
large, unbiased, spectroscopically complete sample.  Unfortunately,
this sample does not exist.  Instead, previous studies have typically
simulated large samples of SNe with the simulated sample properties
matching those believed to be representative of a particular survey
\citep[e.g.,][]{Kessler10}.  This approach is reasonable when
generating samples of SN light curves since there is a significant
amount of light-curve data available and sufficient understanding of
the relative rates of various SN (sub)types and their luminosity
functions \citep{Li11:rate2}.

However, simulations are not necessarily appropriate when looking at
host-galaxy properties of SN samples.  Some observables have not been
examined in detail, and the correlations between properties are not
well understood.  For the purposes of this examination, there is a
large, almost spectroscopically complete sample of SNe that is
relatively free of bias: the Lick Observatory Supernova Search (LOSS)
sample \citep{Leaman11}.

LOSS is a SN search that has run for over a decade monitoring nearby
galaxies with a cadence of a few nights to a couple of weeks.  The
``full'' LOSS sample contains 929 SNe, while the ``optimal'' LOSS
sample contains 726 SNe where 98.3\% of the SNe have a spectroscopic
classification \citep{Leaman11}.  The LOSS detection efficiency is
very high (\about 90\%) with the vast majority of missed objects being
in the nuclear regions of bright, compact galaxies \citep{Leaman11}.
The three major biases for the sample are (1) the missed objects in
nuclear regions, (2) that luminous galaxies are over-represented the
sample, an effect that increases with distance, and (3) that the
Hubble type distribution changes towards earlier galaxy types with
distance.  However, those biases can be somewhat mitigated and do not
affect certain measurements.

After constructing SN luminosity functions for each subtype,
\citet{Li11:rate2} determined that the LOSS sample is relatively
complete to a distance of 80 and 60~Mpc for SNe~Ia and core-collapse
SNe, respectively.  Within 60~Mpc, the $K$-band luminosity function of
the LOSS galaxy sample matches that of a complete sample for galaxies
with $M_{K} < -23$~mag \citep{Leaman11}.  At fainter magnitudes, the
LOSS sample is incomplete.  Similarly, the average $K$-band luminosity
for E and Scd galaxies in the LOSS sample increases by a factor of 4
and 20 from 15 to 175~Mpc, respectively (while the average galaxy
increases by a factor of \about 2 between 15 and 60~Mpc regardless of
Hubble type; \citealt{Leaman11}).

We will examine two subsamples of the LOSS sample.  The ``Full''
sample is nearly equivalent to the ``full'' LOSS sample as defined by
\citet{Leaman11}.  We add classifications for 3 SNe in this sample.
SN~2000cc was observed by \citet{Aldering00:00ca}, who noted that it
had a featureless spectrum consistent with a blackbody.  Although that
is not a definitive classification, it is consistent with a
core-collapse SN and inconsistent with a SN~Ia.  We also classify
SN~2000ft as a SN~II.  This SN has no optical spectrum, but its radio
light curve is consistent with a SN~II \citep{Alberdi06}.  Finally,
\citet{Blondin12} classified SN~2004cu as a SN~Ia.

To generate the ``Full'' sample, we remove 24 SNe from the ``full''
LOSS sample.  Of these 24, 6 are similar to SN~2005E
\citep{Perets10:05e} and 7 are SNe~Iax \citep{Foley13:iax}.  Although
there is evidence that these SNe are peculiar thermonuclear SNe
\citep[e.g.,][]{Li03:02cx, Foley09:08ha, Foley10:08ha, Foley10:08ge,
  Foley13:iax, Perets10:05e}, there is still some controversy
\citep[e.g.,][]{Valenti09}; as a result, we remove these SNe from this
analysis.  We also remove SN~2008J, which appears to be a SN~Ia
interacting with circumstellar hydrogen \citep{Taddia12}.

After all alterations, the ``Full'' sample has 905 SNe, of which 368
are SNe~Ia and 537 are core-collapse SNe (137 SNe~Ibc and 400 SNe~II).

In Section~\ref{s:tests}, we will examine SN samples from the Sloan
Digital Sky Survey (SDSS) SN survey \citep{Frieman08} and Palomar
Transient Factory \citep[PTF;][]{Law09}.  Both surveys are large-area
untargeted surveys; the SN samples are not biased to those in luminous
galaxies.

The SDSS SN survey was performed over three seasons with the SDSS
telescope.  Spectroscopic follow-up was performed with a variety of
telescopes \citep{Zheng08, Konishi11:subaru, Ostman11, Foley12:sdss}.
The first cosmological results based on a spectroscopic sample were
presented by \citet{Kessler09:cosmo}.  A photometric sample was
presented by \citet{Sako11}, and a cosmological analysis of a
photometric SN~Ia sample was performed by \citet{Campbell13}.

PTF has been running a transient survey since 2009 using the 48-inch
telescope at Palomar Observatory.  Although there has not been an
official spectroscopic data release yet, PTF has publicly announced
several hundred spectroscopically classified SNe.

Despite being untargeted surveys, neither SDSS nor PTF are close to
being spectroscopically complete.  Spectroscopically complete
subsamples are much smaller than LOSS.  We therefore choose to focus
on the LOSS sample initially and test our method with the SDSS and PTF
samples.

\subsection{Host-galaxy Observables}

The simplest, although also the least quantitative, metric for
determining bulk host-galaxy properties is the Hubble type.  The
Hubble types for the LOSS galaxies have been determined in a
consistent way and presented by \citet{Leaman11}.  Similarly, the
Galaxy Zoo project has determined visual morphological classifications
for a large number of SDSS galaxies \citep{Lintott11}.  They use the
individual classifications of many volunteers to determine a
probability that a galaxy has an elliptical or spiral morphology.  The
probabilities take into account biases associated with redshift.

In addition to Hubble type, one can easily measure the color of the
host galaxy.  Particular colors correlate well with star-formation
rate, and should therefore correlate with the types of SNe produced.
\citet{Leaman11} present $B$, $B_{0}$, and $K$ band measurements for
the LOSS galaxy sample, where $B_{0}$ is the $B$ magnitude corrected
for Galactic extinction, internal extinction, and $K$-corrections.
Since the $B$ band straddles the 4000~\AA\ break, the $B_{0} - K$
color is a reasonable proxy for the star formation rate.

Galaxy morphology is correlated with both color and luminosity.  More
luminous galaxies tend to be ellipticals, gas poor, and lack recent
star formation.  One can generally cleanly separate star-forming and
passive galaxies using a color-magnitude diagram, with both dimensions
providing information.  We will also examine SN populations as a
function of host-galaxy luminosity.  Specifically, we examine $M_{K}$,
which is highly correlated with stellar mass.

Since core-collapse SNe are associated with star-forming regions
within a galaxy, while SNe~Ia are not, we will examine the proximity
of SN locations to bright regions of the host galaxy.  We use the
``pixel-based'' method of \citet{Fruchter06}, which compares SN
locations to the intensity map of a galaxy, with the brightest pixels
corresponding to a value of 1 and the faintest pixels corresponding to
a value of 0.  We use the values provided by \citet{Kelly08}, which
cover a subsample of the LOSS sample.  We refer to this derived
quantity as the ``pixel rank.''

We also examine the offset of the SN relative to the nucleus.  Both
the underlying stellar population and the progenitor metallicity
should correlate with the offset.  For this measurement, we use the
effective offset, $R$, of \citet{Sullivan06}, which is a dimensionless
parameter describing the separation of the SN from its host galaxy.  A
value of $R = 3$ corresponds roughly to the isophotal limit of the
galaxy.

\section{Galaxy Properties of the LOSS SN Sample}\label{s:galaxy}

\begin{figure*}
\begin{center}
\epsscale{0.55}
\rotatebox{90}{
\plotone{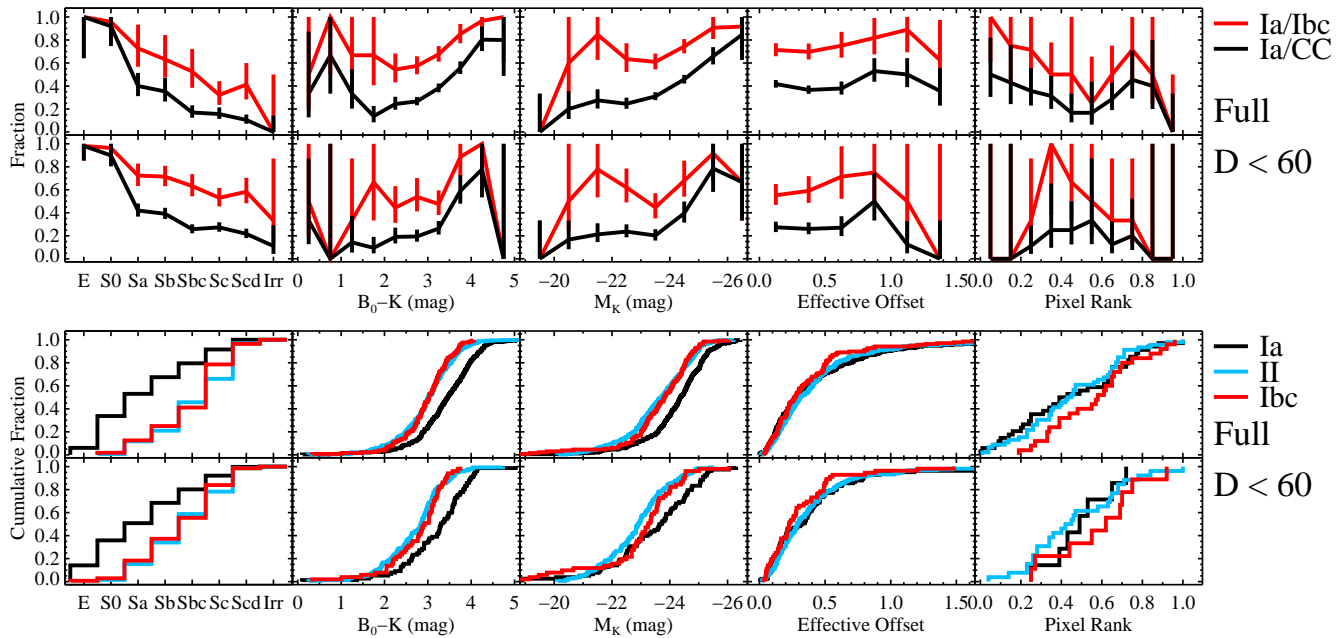}}
\caption{Top panels display the fraction of SNe~Ia in the LOSS sample
  (black) and the subset of SNe~I (red) as a function of host-galaxy
  parameters.  The bottom panels display the CDFs for each host-galaxy
  observable for SNe~Ia (black), SNe~II (blue), and SNe~Ibc (red).
  The first and third (second and last) rows display the results for
  the Full ($D < 60$~Mpc) LOSS sample.  The host-galaxy parameters are
  from left to right, morphology, color, absolute magnitude, effective
  offset, and pixel rank.}\label{f:frac}
\end{center}
\end{figure*}

We now examine how host-galaxy properties can predict SN types in the
LOSS sample.  Figure~\ref{f:frac} displays the fraction of SNe~Ia in
the LOSS sample, and the subset of SNe~I, as a function of host galaxy
property (morphology, color, luminosity, effective offset, and pixel
ranking, respectively).  Figure~\ref{f:frac} also shows the cumulative
distribution functions (CDFs) for SNe~Ia, SNe~II, and SNe~Ibc for each
host-galaxy property.

Consistently, we see that SNe~Ia are more frequently found in galaxies
with properties consistent with older populations than that of the
core-collapse comparison sample.  Specifically, the fraction of SNe
that are of Type~Ia with early-type, red, or luminous host galaxies is
larger than the fraction with late-type, blue, or faint host galaxies.
And therefore, the probability that a particular SN is a SN~Ia is
higher if its host galaxy is a luminous, red, early-type galaxy.  For
instance, 98\% of all SNe with elliptical host galaxies are SNe~Ia,
while only 10\% of all SNe with irregular host galaxies are SNe~Ia.
Clearly, host galaxy information can be useful for classifying SNe~Ia
with no additional information.

This trend continues even within galaxies, where SNe~Ia tend to be
found with larger offsets and in fainter regions of the galaxy than
core-collapse SNe.  Because of the small number of LOSS SNe with
pixel-ranking data, the uncertainties are especially large for that
property.  This metric should be re-examined when more data becomes
available.

Figure~\ref{f:frac} displays results for both the ``Full'' LOSS sample
and the volume-limited LOSS sample.  There are no significant
differences between the samples, and most importantly, the fractions
are consistent for the same bins.  This indicates that whatever biases
the larger LOSS sample has, they have little effect on the fraction of
SNe~Ia from host galaxies that are very similar in one of these
properties.  This result is especially important for transferring
results to other surveys where galaxy population will not be the same
as the LOSS survey.

\section{Galsnid}\label{s:galsnid}

Using the data above, we can create a metric for determining the
probability that a given SN is of Type~Ia.  Specifically, this
probability can be expressed using Bayes' Theorem.  Here, we consider
the case where we only wish to distinguish between two choices, `Ia'
and `Core-collapse' (`CC').  That is, from a classification point of
view, we consider all SNe to have a type, $T \in \{{\rm Ia, CC}\}$.
For a given observable, $D_{i}$, we estimate the probability that a SN
is of Type Ia, $P({\rm Ia} | D_{i})$.  We seek to compute the
probability that a SN is of a given type given multiple observables,
\begin{equation}
  P({\rm Ia} | \boldsymbol{D}),
\end{equation}
where $\boldsymbol{D}$ is the vector of its host-galaxy data.  Since
we are only considering two classes, we have
\begin{equation}
  1 - P({\rm Ia} | \boldsymbol{D}) = P({\rm CC} | \boldsymbol{D}). \label{e:eq1}
\end{equation}
and
\begin{equation}
  1 - P({\rm Ia}) = P({\rm CC}), \label{e:eq2}
\end{equation}
where $P({\rm T})$ is the overall probability of a
given SN in the sample is of Type $T$ (the prior).

Bayes' Theorem is
\begin{equation}
  P({\rm Ia} | \boldsymbol{D}) = k^{-1} P(\boldsymbol{D} | {\rm Ia}) P({\rm Ia}),
\end{equation}
where $k^{-1}$ is a normalization factor depending on $\boldsymbol{D}$
set by requiring the class probabilities to add to unity and
$P(\boldsymbol{D} | {\rm Ia})$ is the probability density of a set of
observables given that the object is a SN Ia.  The likelihood
$P(\boldsymbol{D} | {\rm Ia})$ is difficult to model directly since
$\boldsymbol{D}$ is multi-dimensional.  It is convenient to neglect
the correlations among the galaxy observables and make the
approximation that their joint probability factors as the product of
the individual one-dimensional likelihoods of each galaxy property.
We can do this by invoking the Naive Bayes assumption that the data
are conditionally independent given the class\footnote{This assumption
  is typically not true; see discussion of this limitation in
  Section~\ref{ss:improve}.}, which gives us
\begin{equation}
  P({\rm Ia} | \boldsymbol{D}) = k^{-1} P({\rm Ia}) \prod_{i = 1}^{n} P(D_{i} | {\rm Ia}), \label{e:galsnid}
\end{equation}
where $D_{i}$ are the individual $n$ observables.

From Equations~\ref{e:eq1}, \ref{e:eq2}, and \ref{e:galsnid},
\begin{equation}
  k = P({\rm Ia}) \prod_{i = 1}^{n} P(D_{i} | {\rm Ia}) + (1 - P({\rm Ia}))\prod_{i = 1}^{n} P(D_{i} | {\rm CC}).
\end{equation}

The underlying population of the SNe and host galaxies are somewhat
important in the determination of the probability.  As an example, we
consider a single observable.  In that case, we have
\begin{equation}
  P({\rm Ia} | \boldsymbol{D}) = k^{-1} P({\rm Ia}) P(D_{1} | {\rm Ia}).
\end{equation}
With some algebraic manipulation, we find that
\begin{equation}
  P({\rm Ia} | \boldsymbol{D}) = \frac{f_{{\rm Ia}/{\rm CC}, x}}{1 + f_{{\rm Ia}/{\rm CC}, x}}
\end{equation}
where $f_{{\rm Ia}/{\rm CC}, x}$ is the odds ratio of SN~Ia to
core-collapse SN for a particular value of $D_{1}$ (the relative
fraction of SNe~Ia to core-collapse SNe with $D_{1} = x$).  This
implies that biased samples can still be useful for determining
probabilities for {\it all} other samples, both biased and unbiased,
as long as the samples retain the same relative fraction of SNe~Ia and
core-collapse SNe for a particular value of each observable.
Specifically, the known biases of the LOSS sample should not affect
our ability to apply its results to other low-redshift samples.
However, using the LOSS sample for high-redshift SNe where, for
example, we know that the relative fractions of SNe~Ia and
core-collapse SNe is different in spirals, will bias the results
somewhat.

From the LOSS data, we have determined $P(D_{i} | {\rm Ia})$ and
$P(D_{i} | {\rm CC})$ for all relevant values of each observable; this
is simply the fraction of SN Ia (or core-collapse SN) host galaxies
that have a particular galaxy observable, $D_{i}$.  We present these
data in Table~\ref{t:prop}.  Since some bins do not contain many SNe,
the uncertainties can be large for some values of particular
parameters.  To avoid potential biases associated with large
statistical uncertainties, we perform a Monte Carlo simulation for
each SN where we determine $\tilde{P}(D_{i} | T)$, a single
realization of the probability for variable $i$ using $P(D_{i} | T)$
and its uncertainty.  This Monte Carlo is performed for all
observables simultaneously, resulting in several realizations of the
overall probability that a given SN is of Type~Ia, $\tilde{P} ({\rm
  Ia} | \boldsymbol{D})$.  From the Monte Carlo simulation, we have a
distribution of posterior probabilities that each SN is of Type Ia.
We then assign the final probability, $P ({\rm Ia} | \boldsymbol{D})$
to be the median value of the distribution of $\tilde{P} ({\rm Ia} |
\boldsymbol{D})$.

\begin{deluxetable}{lll}
\tabletypesize{\scriptsize}
\tablewidth{0pt}
\tablecaption{Probability of Host Properties Given Type\label{t:prop}}
\tablehead{
\colhead{Bin} &
\colhead{$P(D_{i} | {\rm Ia})$} &
\colhead{$P(D_{i} | {\rm CC})$}}

\startdata

\multicolumn{3}{c}{Morphology} \\
\tableline
E   & 0.141 \err{0.021}{0.018} & 0.002 \err{0.003}{0.001} \\
S0  & 0.217 \err{0.026}{0.023} & 0.017 \err{0.007}{0.005} \\
Sa  & 0.149 \err{0.022}{0.019} & 0.142 \err{0.017}{0.015} \\
Sb  & 0.177 \err{0.023}{0.021} & 0.188 \err{0.020}{0.018} \\
Sbc & 0.117 \err{0.019}{0.017} & 0.231 \err{0.022}{0.020} \\
Sc  & 0.120 \err{0.019}{0.017} & 0.218 \err{0.021}{0.019} \\
Scd & 0.076 \err{0.016}{0.013} & 0.188 \err{0.020}{0.018} \\
Irr & 0.003 \err{0.004}{0.002} & 0.015 \err{0.006}{0.004} \\
\tableline
\multicolumn{3}{c}{$B_{0}-K$} \\
\tableline
0.0  -- 1.75 & 0.026 \err{0.010}{0.007} & 0.044 \err{0.010}{0.008} \\
1.75 -- 2.25 & 0.023 \err{0.010}{0.007} & 0.075 \err{0.013}{0.011} \\
2.25 -- 2.5  & 0.037 \err{0.012}{0.009} & 0.069 \err{0.013}{0.011} \\
2.5  -- 2.75 & 0.043 \err{0.013}{0.010} & 0.115 \err{0.016}{0.014} \\
2.75 -- 3.0  & 0.111 \err{0.019}{0.016} & 0.181 \err{0.020}{0.018} \\
3.0  -- 3.25 & 0.131 \err{0.021}{0.018} & 0.185 \err{0.020}{0.018} \\
3.25 -- 3.5  & 0.154 \err{0.022}{0.020} & 0.137 \err{0.018}{0.016} \\
3.5  -- 3.75 & 0.125 \err{0.020}{0.018} & 0.115 \err{0.016}{0.014} \\
3.75 -- 4.0  & 0.168 \err{0.023}{0.021} & 0.048 \err{0.011}{0.009} \\
4.0  -- 4.25 & 0.103 \err{0.019}{0.016} & 0.022 \err{0.008}{0.006} \\
4.25 -- 6.25 & 0.080 \err{0.017}{0.014} & 0.010 \err{0.006}{0.004} \\
\tableline
\multicolumn{3}{c}{$M_{K}$} \\
\tableline
$-26.5$ -- $-25.5$ & 0.079 \err{0.016}{0.014} & 0.018 \err{0.007}{0.005} \\
$-25.5$ -- $-25.1$ & 0.108 \err{0.019}{0.016} & 0.029 \err{0.009}{0.007} \\
$-25.1$ -- $-24.7$ & 0.164 \err{0.023}{0.020} & 0.125 \err{0.017}{0.015} \\
$-24.7$ -- $-24.3$ & 0.190 \err{0.025}{0.022} & 0.146 \err{0.018}{0.016} \\
$-24.3$ -- $-23.9$ & 0.153 \err{0.022}{0.019} & 0.125 \err{0.017}{0.015} \\
$-23.9$ -- $-23.5$ & 0.130 \err{0.021}{0.018} & 0.136 \err{0.017}{0.015} \\
$-23.5$ -- $-23.1$ & 0.040 \err{0.012}{0.009} & 0.133 \err{0.017}{0.015} \\
$-23.1$ -- $-22.7$ & 0.048 \err{0.013}{0.010} & 0.115 \err{0.016}{0.014} \\
$-22.7$ -- $-22.3$ & 0.031 \err{0.011}{0.008} & 0.058 \err{0.012}{0.010} \\
$-22.3$ -- $-21.5$ & 0.037 \err{0.012}{0.009} & 0.060 \err{0.012}{0.010} \\
$-21.5$ -- $-17.1$ & 0.020 \err{0.009}{0.006} & 0.055 \err{0.011}{0.009} \\
\tableline
\multicolumn{3}{c}{Effective Offset} \\
\tableline
0.0  -- 0.05 & 0.043 \err{0.012}{0.010} & 0.032 \err{0.009}{0.007} \\
0.05 -- 0.1  & 0.098 \err{0.018}{0.015} & 0.086 \err{0.014}{0.012} \\
0.1  -- 0.15 & 0.130 \err{0.020}{0.018} & 0.091 \err{0.014}{0.012} \\
0.15 -- 0.2  & 0.090 \err{0.017}{0.014} & 0.091 \err{0.014}{0.012} \\
0.2  -- 0.25 & 0.071 \err{0.015}{0.013} & 0.114 \err{0.016}{0.014} \\
0.25 -- 0.3  & 0.057 \err{0.014}{0.011} & 0.084 \err{0.013}{0.012} \\
0.3  -- 0.35 & 0.068 \err{0.015}{0.012} & 0.058 \err{0.011}{0.009} \\
0.35 -- 0.4  & 0.060 \err{0.014}{0.011} & 0.054 \err{0.011}{0.009} \\
0.4  -- 0.45 & 0.038 \err{0.012}{0.009} & 0.067 \err{0.012}{0.010} \\
0.45 -- 0.5  & 0.052 \err{0.013}{0.011} & 0.063 \err{0.012}{0.010} \\
0.5  -- 0.6  & 0.060 \err{0.014}{0.011} & 0.089 \err{0.014}{0.012} \\
0.6  -- 0.75 & 0.062 \err{0.014}{0.012} & 0.048 \err{0.010}{0.009} \\
0.75 -- 1.0  & 0.073 \err{0.016}{0.013} & 0.045 \err{0.010}{0.008} \\
1.0  -- 1.4  & 0.052 \err{0.013}{0.011} & 0.041 \err{0.010}{0.008} \\
1.4  -- 5.25 & 0.046 \err{0.013}{0.010} & 0.037 \err{0.009}{0.007} \\
\tableline
\multicolumn{3}{c}{Pixel Rank} \\
\tableline
0.0 -- 0.2 & 0.206 \err{0.094}{0.064} & 0.113 \err{0.047}{0.033} \\
0.2 -- 0.4 & 0.294 \err{0.109}{0.079} & 0.282 \err{0.070}{0.056} \\
0.4 -- 0.6 & 0.088 \err{0.068}{0.038} & 0.211 \err{0.062}{0.048} \\
0.6 -- 0.8 & 0.324 \err{0.113}{0.084} & 0.296 \err{0.072}{0.058} \\
0.8 -- 1.0 & 0.088 \err{0.068}{0.038} & 0.099 \err{0.045}{0.031}

\enddata

\end{deluxetable}

Following the convention of {\it psnid} \citep[photometric supernova
identification;][]{Sako11}, we call this procedure {\it galsnid}
(galaxy-property supernova identification).  For the rest of the
manuscript, we define the posterior probability from {\it galsnid} as
$p \equiv P({\rm Ia} | \boldsymbol{D})$.

\begin{figure}
\begin{center}
\epsscale{1.0}
\rotatebox{90}{
\plotone{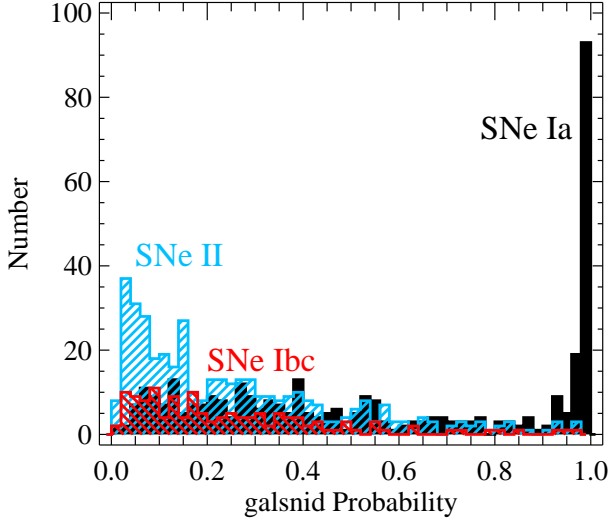}}
\caption{Histogram of {\it galsnid} probability for different
  spectroscopic SN classes in the LOSS sample.  The filled black, red,
  and blue histograms represent SNe~Ia, Ibc, and II,
  respectively.}\label{f:hist}
\end{center}
\end{figure}

Having performed this procedure for the Full LOSS sample, we arrive
with a distribution of probabilities from 0 to 1.  We display the
results in Figure~\ref{f:hist}.  In this Figure, we present histograms
for the probability that a SN is of Type~Ia for the spectroscopically
confirmed subsets of SNe~Ia, II, and Ibc.  For the LOSS sample, 30\%
have $p > 0.5$; of those 71\% are SNe~Ia.  This compares favorably to
the prior of $P({\rm Ia}) = 0.41$.  For the same sample, 21\% have $p
< 0.1$; 84\% of which are core-collapse SNe.

\begin{figure}
\begin{center}
\epsscale{1.0}
\rotatebox{90}{
\plotone{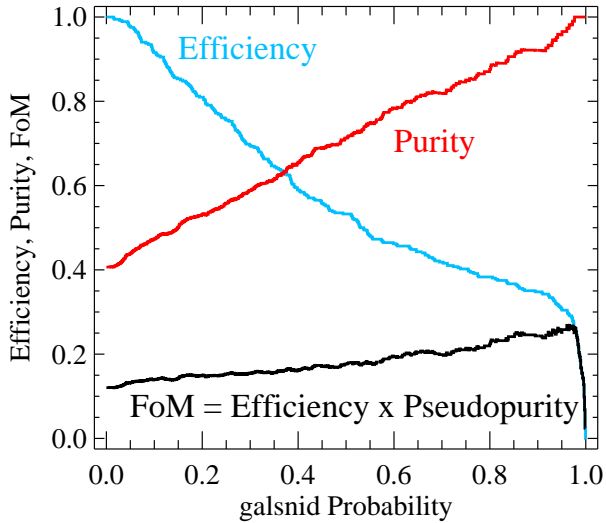}}
\caption{Efficiency, purity, and FoM (blue, red, and black curves,
  respectively) for subsamples of the LOSS sample defined by a
  particular {\it galsnid} probability or larger.  The FoM peaks at $p
  = 0.97$ at a value that is 2.23 times larger than the FoM for the
  entire sample.}\label{f:fom}
\end{center}
\end{figure}

Again, this information can both be used by itself and in combination
with SN photometry for classification.  We test the utility of using
only the {\it galsnid} method with the FoM defined in
Section~\ref{s:fom}.  Figure~\ref{f:fom} presents the efficiency, the
purity ($W^{\rm False}_{\rm Ia} = 1$), and the FoM (assuming $W^{\rm
  False}_{\rm Ia} = 5$) for subsamples including only objects with
{\it galsnid} probability $p$ greater than a threshold value.  The FoM
peaks at $p = 0.97$ at a value of 0.269.  The full sample has a FoM of
0.121, so implementing {\it galsnid} improves the FoM by a factor of
2.23.  As a comparison, \citet{Campbell13} performed {\it psnid} on a
simulated sample of SNe~Ia and obtained an improvement of 2.60.

\section{Tests}\label{s:tests}

In this section, we perform a variety of tests on the {\it galsnid}
method for classifying SNe.  Specifically, we examine the reliability
of the method, test the importance of each galaxy observable for
classification, and apply the method to additional SN samples.

\subsection{Cross-validation}

Having shown that host-galaxy information is useful for SN
classification, we now test the robustness of the above results.
Specifically, we cross-validate the method again using the LOSS
sample.  We split the sample in half, placing every-other SN (to
mitigate possible biases in the SN search or classification with time)
as the training and comparison samples.  Using the ``evenly-indexed''
sample as the training set, we find the {\it galsnid} probability
which results in the highest FoM for the training set, which we
consider the threshold value above which a SN will enter our final
sample.  We then apply the probabilities and this threshold {\it
  galsnid} probability found from the training set to the
``oddly-indexed'' sample and determine the efficiency and
pseudo-purity of the sample.  Doing this, we find that the FoM
improves by a factor of 1.4 compared to not using the {\it galsnid}
procedure.  Performing the same procedure but switching the training
and testing samples, we find that the FoM improves by a factor of 2.4.
Therefore, the method appears to be robust within a given sample,
although clearly the amount of improvement depends on the training
sample.

\subsection{Importance of Each Observable}

To assess the importance of each host-galaxy observable for
classification, we first re-analyze the data and computed {\it
  galsnid} probabilities using a single observable at a time.  We then
compute the {\it galsnid} probabilities using all observables, but
excluding a single observable at a time.  The summary of the results
are presented in Table~\ref{t:imp}, where we list the peak FoM, the
improvement factor over the baseline FoM, and the difference in the
median {\it galsnid} probability for the spectroscopically confirmed
SNe~Ia and core-collapse SN classes.  The latter is a measure of the
difference of distribution of {\it galsnid} probabilities for
different spectroscopic classes, and thus an additional (and
different) indication of the importance of the observable beyond the
improvement in the FoM.

\begin{deluxetable*}{lcccccc}
\tabletypesize{\scriptsize}
\tablewidth{0pt}
\tablecaption{Importance of Each Observable\label{t:imp}}
\tablehead{
&
\multicolumn{3}{c}{Exclusively Using Observable} &
\multicolumn{3}{c}{Excluding Observable} \\
&
\multicolumn{3}{c}{---------------------------------------------} &
\multicolumn{3}{c}{---------------------------------------------} \\
\colhead{} &
\colhead{Peak} &
\colhead{Improvement} &
\colhead{Difference} &
\colhead{Peak} &
\colhead{Improvement} &
\colhead{Difference} \\
\colhead{Observable} &
\colhead{FoM} &
\colhead{Factor} &
\colhead{in Medians} &
\colhead{FoM} &
\colhead{Factor} &
\colhead{in Medians}}

\startdata

Baseline\tablenotemark{a}              & 0.121 & N/A  & N/A  & 0.121 & N/A  & N/A  \\
Using All Galaxy Data\tablenotemark{b} & 0.269 & 2.23 & 0.34 & 0.269 & 2.23 & 0.34 \\
Morphology                             & 0.262 & 2.18 & 0.15 & 0.157 & 1.30 & 0.22 \\
Color                                  & 0.128 & 1.06 & 0.10 & 0.273 & 2.26 & 0.26 \\
Luminosity                             & 0.135 & 1.12 & 0.07 & 0.273 & 2.27 & 0.24 \\
Effective Offset                       & 0.122 & 1.02 & 0.03 & 0.261 & 2.16 & 0.30 \\
Pixel Rank                             & 0.123 & 1.02 & 0.00 & 0.269 & 2.23 & 0.33

\enddata

\tablenotetext{a}{The ``Baseline'' category classifies the entire SN
  sample as SN~Ia without using host-galaxy information.}

\tablenotetext{b}{This category is for the nominal {\it galsnid}
  procedure, as defined in Section~\ref{s:galsnid}, using all
  host-galaxy data.}

\end{deluxetable*}

Unsurprisingly, the pixel ranking data was not particularly useful,
and excluding it made no significant difference in the results.  The
vast majority of SNe in the sample do not have pixel ranking data, and
thus it only has the ability to affect a small number of objects.
Additionally, pixel ranking does not appear to be as discriminating as
other observables.

The color and luminosity are both somewhat important.  The median {\it
  galsnid} probability when just using color (luminosity) was 0.43
(0.47) and 0.33 (0.40) for SNe~Ia and core-collapse SNe, respectively.
However, just using color or luminosity results in only a modest
improvement in the peak FoM with ratios of 1.06 and 1.12,
respectively.  Using both quantities together (but excluding all other
observables) results in a maximum FoM improvement ratio of 1.17.

Removing either color and luminosity results in only modest changes in
the maximum FoM from 0.269 to 0.273 (a net increase).  We do not
consider this change in the FoM significant.  However, removing these
data results in more smearing of the populations with the difference
in the median {\it galsnid} probabilities for SNe~Ia and core-collapse
SNe decreasing from 0.34 to 0.26 and 0.24, respectively.  Removing
both color and luminosity continues this trend with a difference in
medians for the two populations of only 0.16.  Therefore, although
color and luminosity do not significantly affect the peak FoM
presented here, they could be particularly important for other
applications or different FoMs.

Using only offset information results in no significant improvement in maximum
FoM, although it is slightly helpful with classification; the median
{\it galsnid} values for the SN Ia and core-collapse populations are
0.44 and 0.41, respectively.  However, removing the relative offset
decreases the maximum FoM to 0.261.  This is a somewhat surprising
result and may not be significant.

By far, the most important parameter is morphology.  Morphology alone
results in a maximum FoM of 0.262, a factor of 2.18 improvement over
not using any host-galaxy information.  Removing morphology
information decreases the maximum FoM to 0.157.  Without these data,
the maximum improvement is only a factor of 1.30 over not using any
host-galaxy data.  Nonetheless, {\it galsnid} is still effective
without morphology.

\subsection{SDSS}\label{ss:sdss}

We also examined the largest photometry-selected SN~Ia sample: the
SDSS-II SN survey compilation \citep{Sako11, Campbell13}.  This
sample, which we call the ``SDSS sample,'' was taken from the SDSS-II
SN survey and various cuts were made based on the photometric
properties of the SNe to determine a relatively pure subsample of
SNe~Ia (see \citealt{Campbell13} for details).  This sample is a
subset of the full photometric-only sample of SNe from SDSS-II
\citep{Sako11}.  All SNe in the SDSS sample have host-galaxy
redshifts.

Using simulations, \citet{Campbell13} showed that the SDSS sample
should have an efficiency of 71\% and a contamination of 4\%.  This
sample only includes SNe photometrically classified as Type Ia.

Using the SDSS imaging data of the host galaxies, we apply the {\it
  galsnid} procedure to the SDSS sample.  This sample is already
supposedly quite pure, and an ideal method would provide a criteria to
sift out the 4\% contamination without a large loss of true SNe~Ia.
Of course this is not a perfect test of a given method since increased
efficiency with minimal decrease in purity is also a net gain.
Nonetheless, a qualitative assessment can be made.

For this test, we focused on the SNe with $z < 0.2$.  This subsample
is likely to have some contamination from core-collapse SNe; Malmquist
bias will remove many low-luminosity core-collapse SNe from the
higher-redshift sample.  The efficiency for this subsample is also
expected to be higher, providing a more representative SN~Ia sample.
For these redshifts, one may also expect that the fractions of
different SN types for a given galaxy property have not evolved much.
For the SDSS sample, the SN parameters (light-curve shape and color)
do not evolve much for this redshift range.  Additionally, a
significant number of SNe in this sample are spectroscopically
confirmed as SNe~Ia.  Finally, \citet{Campbell13} shows that
simulations predict several large Hubble-diagram outliers from
core-collapse SNe at $z < 0.2$ that remain in the sample.  Although
there is no direct evidence of that contamination, there are also
several Hubble-diagram outliers at $z < 0.2$ in the data as well.
Understanding this potential contamination and potentially identifying
a solution would be useful.  This subsample contains 143 SNe, but only
131 have matches for the listed galaxy ID in DR8/9 \citep{Ahn12,
  Aihara12}.  We require host-galaxy photometry and a host-galaxy
redshift for this analysis.

The overall fraction of SNe~Ia in the Full and volume-limited LOSS
sample is 40\% and 27\%, respectively.  \citet{Bazin09} found that
only 18\% of SNe at $z < 0.4$ in the Supernova Legacy Survey were
SNe~Ia.  However, using the volumetric rates as a function of redshift
from \citet{Dilday08} and \citet{Bazin09} (for SNe~Ia and
core-collapse SNe, respectively), we find that the SN~Ia fraction
should be 22\% at $z = 0.15$.  Of course this is the {\it volumetric}
fraction.  SDSS is relatively complete to $z = 0.2$, but still suffers
from some Malmquist bias.  The magnitude-limited SN~Ia fraction in the
LOSS sample was 79\% \citep{Li11:rate2}.  For the SDSS sample, we take
an intermediate value of 50\%.  This number essentially provides a
normalization for the probability and does not affect relative
results.

Taking the DR8 imaging data, we determined the \ugriz\ magnitudes for
each host galaxy.  Cross-checking with earlier data releases, we
verified that no measurements were significantly affected by SN light.
Using the {\it kcorrect} routine \citep[version 4\_1\_4;][]{Blanton03,
  Blanton07}, we calculated rest-frame $B$ and $K$ magnitudes.  This
method extrapolates galaxy templates into the NIR to estimate the $K$
magnitudes, but since these galaxy have 5-band photometry, including
$z$ band, this extrapolation should be relatively robust.  We tested
this by comparing the observer-frame $z$ and extrapolated
observer-frame $K$ photometry; the two values were highly correlated.
From the derived photometry, we were able to measure $B-K$ and $M_{K}$
for each SDSS host galaxy.

We also used morphological classifications from the Galaxy Zoo survey
\citep{Lintott11}. Only 33 of the 131 SN host galaxies have
morphological classifications from Galaxy Zoo.  Using the LOSS data,
we redetermined the probabilities of a SN being of Type Ia when only
using the coarse bins of ``elliptical'' and ``spiral.''  These
probabilities are used for the SDSS galaxies with morphology
information.

Since the effective offset and pixel ranking are not particularly
effective at classifying SNe, we do not use those measurements.

For comparison, we also chose 10,000 random galaxies in the SDSS-II SN
survey footprint with $z < 0.2$.  We expect that these galaxies will
have average properties that are different from both the average SN~Ia
host galaxy and the average SN host galaxy.  As a result, performing
the {\it galsnid} procedure on these galaxies should provide a
baseline for any potential improvement.

\begin{figure}
\begin{center}
\epsscale{1.0}
\rotatebox{90}{
\plotone{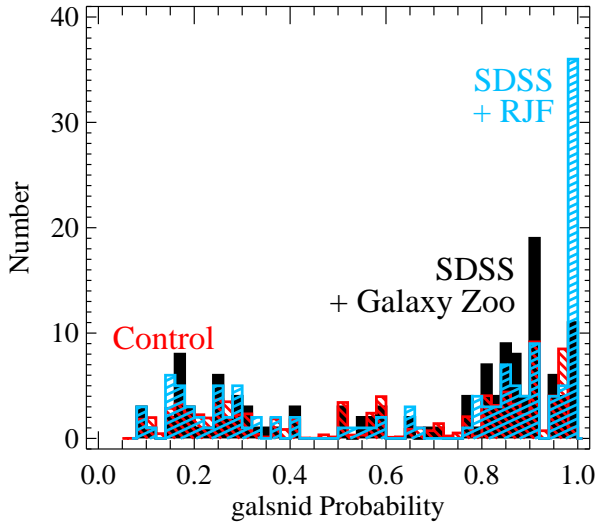}}
\caption{Histogram of {\it galsnid} probability for the SDSS (black
  and blue; 131 galaxies) and control (red; 10,000 galaxies) samples.
  The black and blue histograms show the distributions of
  probabilities when using morphology classifications from Galaxy Zoo
  and one of the authors (RJF), respectively.  The control sample
  histogram has been scaled by a factor of 0.093 to roughly match the
  SDSS histograms.}\label{f:sdss_hist}
\end{center}
\end{figure}

The {\it galsnid} probabilities are shown for these two samples in
Figure~\ref{f:sdss_hist}.  We find that the median {\it galsnid}
probabilities for the SDSS sample and the random comparison sample are
0.82 and 0.66, respectively.  The SDSS host-galaxy sample is more
likely to host SNe~Ia (relative to core-collapse SNe) than the random
sample.  This is further validation of the {\it galsnid} procedure.
However, the probabilities are more evenly distributed than for the
LOSS sample.  We attribute this mainly to the lack of morphology
information for the majority of the sample.

We examined SDSS images for each host galaxy in the SDSS sample and
one of us (RJF) visually classified their morphology as elliptical or
spiral.  We were able to assign a morphology to 87 of the 131
galaxies, of which 36 and 51 were ellipticals and spirals,
respectively.  These classifications are not as robust as the Galaxy
Zoo measurements, but are helpful for assessing how morphology data
can improve our classifications.  After including these new morphology
measurements (and ignoring all Galaxy Zoo measurements), we find that
the median {\it galsnid} probability of the SDSS sample is 0.85,
slightly above the median of the previous analysis.  Moreover, the
number of SNe with $p > 0.97$, the peak of the FoM from the prior
analysis, more than tripled from 11 to 36 SNe.

\begin{figure}
\begin{center}
\epsscale{0.75}
\rotatebox{90}{
\plotone{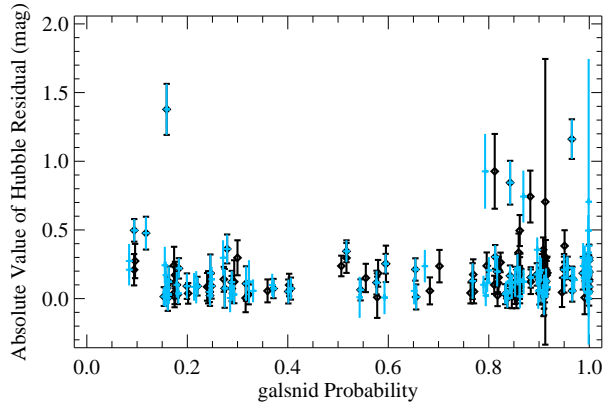}}
\caption{Absolute value of Hubble residuals (assuming the preferred
  \citealt{Campbell13} cosmology) for the SDSS sample as a function of
  {\it galsnid} probability.  The black diamonds (with hats) and blue
  dashes (without hats) represent probabilities using morphology
  classifications from Galaxy and the author (RJF),
  respectively.}\label{f:galsnid_resid}
\end{center}
\end{figure}

Using the best-fit cosmology of \citet{Campbell13}, we are also able
to measure the Hubble residual for each SN in the sample.  Notably,
there are 6 (9) SNe with a Hubble residual $>$0.5 (0.4)~mag from zero.
Figure~\ref{f:galsnid_resid} displays the absolute value of the Hubble
residual as a function of {\it galsnid} probability.  Interestingly, 3
of these SNe, including the most discrepant outlier, have $p < 0.2$.
There are only 18 SDSS SNe with $p < 0.2$, and thus the outliers make
up 17\% of the low-probability subset.

Splitting the SDSS sample by $p = 0.85$ (the median value), we can
examine the characteristics of SNe with low/high {\it galsnid}
probability.  The average and median redshifts for the two samples are
nearly identical, with the high-probability subsample being at
slightly higher redshift (by 0.01).

\begin{figure}
\begin{center}
\epsscale{0.75}
\rotatebox{90}{
\plotone{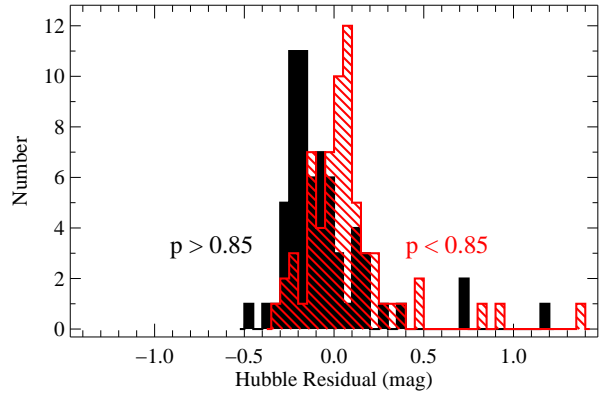}}
\caption{Histograms of Hubble residuals (assuming the preferred
  \citealt{Campbell13} cosmology) for the SDSS sample.  The solid
  black and hashed red histograms represent the samples with $p >
  0.85$ and $p < 0.85$, respectively.}\label{f:sdss_resid}
\end{center}
\end{figure}

Looking at the subsamples in detail, we see a correlation between
Hubble residual (not the absolute value) and {\it galsnid}
probability.  SNe with large {\it galsnid} probability tend to have
negative Hubble residuals, while those with small {\it galsnid}
probability tend to have positive Hubble residuals.  We display these
residuals in Figure~\ref{f:sdss_resid}.  The medians for these
subsamples are $-0.138$ and 0.041~mag, respectively.  The weighted
means are $-0.085 \pm 0.011$ and $0.034 \pm 0.012$.  The difference in
the weighted means are 7.3-$\sigma$ different.  Performing a
Kolmogorav-Smirnov test on the two samples results in a $p$-value of
$4.4 \times 10^{-6}$, indicating that the Hubble residuals are drawn
from different populations.

Considering that most of the SNe in the $z < 0.2$ subsample are
spectroscopically confirmed as Type~Ia, the difference in Hubble
residuals is probably not the result of contamination.  Rather, the
difference is likely related to the known correlation between Hubble
residuals and host-galaxy properties \citep{Kelly10, Lampeitl10:host,
  Sullivan10}.  \citet{Campbell13} chose not to include this
correction.  Since {\it galsnid} probability correlates strongly with
these host-galaxy properties, this is likely the cause of the
difference in Hubble residuals.  However, not accounting for this
effect prevents some analysis of the correlations between Hubble
residuals and {\it galsnid} probability.

Excluding the outlier SNe (Hubble residuals $>$0.5~mag), we fit
Gaussians to the residuals.  Splitting the sample by $p = 0.85$, we
find that the standard deviation of the residuals are 0.128 and
0.114~mag for the subsample with $p \le 0.85$ and $p > 0.85$,
respectively; the subsample with higher {\it galsnid} probability has
smaller scatter.  It is unclear if the difference is the result of
different amounts of contamination in the subsamples or because of the
properties of SNe~Ia in redder, more luminous, earlier galaxies tend
to produce a more standard sample.

The host galaxies of the six outlier SNe (Hubble residual $>$0.5~mag)
are somewhat varied: two large spiral galaxies, an incredibly small
and low-luminosity galaxy, a modest disk galaxy, a small red galaxy
with no signs of star formation, and a small starburst galaxy with a
potential tidal tail.  There is no obvious trend with the host-galaxy
properties investigated here.  Nonetheless, {\it galsnid} provides
some handle on these outliers.

\subsection{PTF}\label{ss:ptf}

To further test the {\it galsnid} method on another independent
sample, we investigate the relatively large sample of publicly
classified SNe from PTF.  PTF is a low-redshift (typically $z < 0.15$)
SN survey that has spectroscopically classified almost 2000 SNe (as of
1 June 2013).  Many of these SNe are publicly announced with
coordinates, redshifts, and classifications.  PTF provides a
relatively large sample of SNe for which we can attempt classification
through host-galaxy properties.

Using the WISeREP database \citep{Yaron12}, we obtained a list of 555
PTF SNe with classifications.  We visually cross-referenced this list
with SDSS images to determine the host galaxy for each SN.  Many SNe
were not in the SDSS footprint, had no obvious host galaxy, or there
was some ambiguity as to which galaxy was the host.  After removing
these objects, 384 SNe remained.  We further restricted the sample to
SNe with host galaxies that have SDSS spectroscopy, leaving a total of
151 SNe.  This sample contains 118 SNe~Ia and 33 core-collapse SNe.
We again match the Galaxy Zoo morphology classifications to this
sample.  For the PTF sample, 131 host galaxies had classifications.

\begin{figure}
\begin{center}
\epsscale{1.0}
\rotatebox{90}{
\plotone{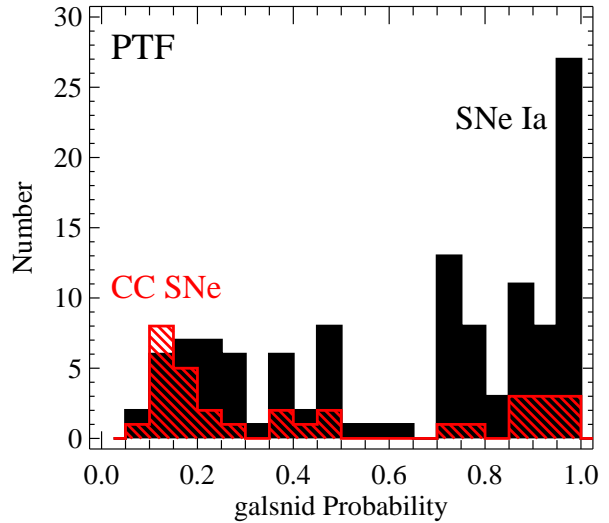}}
\caption{Histogram of {\it galsnid} probability for different
  spectroscopic SN classes in the PTF sample.  The filled black and
  red histograms represent SNe~Ia and core-collapse SNe,
  respectively.}\label{f:ptf_hist}
\end{center}
\end{figure}

Using the same method as described in Section~\ref{ss:sdss}, except
using the LOSS prior on the SN~Ia fraction, we applied the {\it
  galsnid} technique to the PTF sample.  Histograms of the resulting
probabilities are presented in Figure~\ref{f:ptf_hist}.  Again, the
SNe~Ia typically have a much higher {\it galsnid} probability than the
core-collapse SNe.  The median probability for the SNe~Ia and
core-collapse SNe are 0.74 and 0.27, respectively.  This is empirical
proof that simply using the LOSS parameters are useful for classifying
SNe in an untargeted survey.

\section{Discussion}\label{s:disc}

\subsection{Additional Applications}

\begin{figure}
\begin{center}
\epsscale{1.0}
\rotatebox{90}{
\plotone{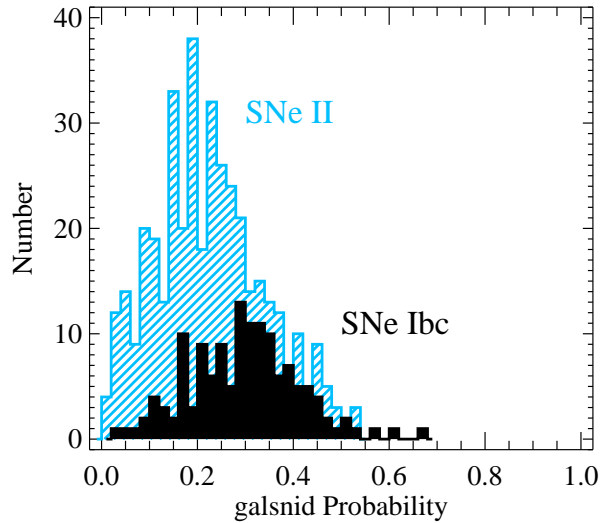}}
\caption{Histogram of {\it galsnid} probability for SNe~Ibc (filled
  black histogram) and SNe~II (dashed blue histogram)
  in the LOSS sample.  The {\it galsnid} probability lists the
  probability that a given SN is of Type Ibc.  The median {\it
    galsnid} probability for all SNe is relatively low to reflect the
  relatively low random chance of a SN being a SN~Ibc (the prior) of
  0.26.}\label{f:cc_hist}
\end{center}
\end{figure}

Although we have focused on separating SNe~Ia from core-collapse SNe,
the {\it galsnid} algorithm can be used for a variety of purposes.  As
an example, we have used the {\it galsnid} algorithm to separate
SNe~Ibc from SNe~II in the LOSS sample (Figure~\ref{f:cc_hist}).
Although the two samples do not separate as cleanly as SNe~Ia from
core-collapse SNe, one can use {\it galsnid} to prioritize SNe for
follow-up.  The method can be particularly useful when a SN is young,
before any light-curve fitting can be performed.

\begin{figure}
\begin{center}
\epsscale{1.0}
\rotatebox{90}{
\plotone{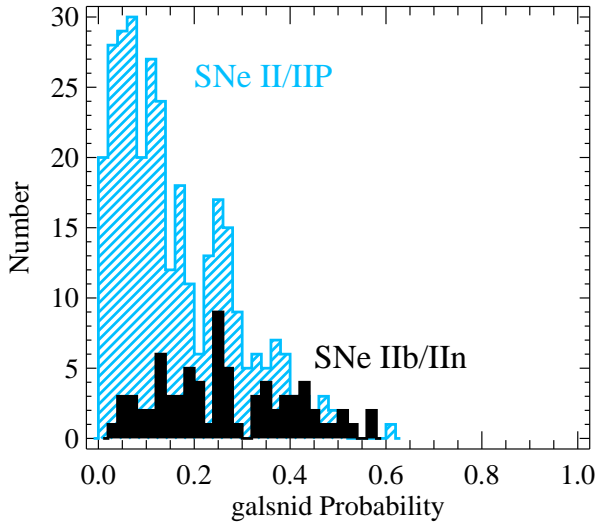}}
\caption{Histogram of {\it galsnid} probability for SNe~IIb/IIn (filled
  black histogram) and SNe~II/IIP (dashed blue histogram)
  in the LOSS sample.  The {\it galsnid} probability lists the
  probability that a given SN is of Type IIb or IIn.  The median {\it
    galsnid} probability for all SNe is relatively low to reflect the
  relatively low random chance of a SN being a SN~IIb/IIn (the prior) of
  0.19.}\label{f:ii_hist}
\end{center}
\end{figure}

Similarly, we can even separate different subclasses of SNe.  As an
example, {\it galsnid} was used to separate ``peculiar'' SNe~II from
``normal'' SNe~II (Figure~\ref{f:ii_hist}).  Specifically, we were
able to separate SNe classified as SNe~IIb or SNe~IIn from those
classified as SNe~IIP or simply SNe~II.  Again, {\it galsnid} is
useful for selecting SNe for follow-up.  In this case, it could be
particularly useful since early spectra of all types of SNe~II can be
relatively featureless, and thus, there could be epochs where the
host-galaxy information is more discriminating than a spectrum.

Another potential application is classifying the small number of
unclassified SNe in the LOSS sample.  For these SNe, we applied the
{\it galsnid} procedure using the LOSS priors and probabilities.
Since these SNe were part of the LOSS sample, it is reasonable to
assume that the absolute probabilities are correct.  That is, a SN with
$p > 0.5$ is likely a SN~Ia.  We list the results in Table~\ref{t:unk}.

\begin{deluxetable}{lcccc}
\tabletypesize{\scriptsize}
\tablewidth{0pt}
\tablecaption{Spectroscopically Unclassified LOSS SNe\label{t:unk}}
\tablehead{
\colhead{SN} &
\colhead{{\it galsnid} $p$} &
\colhead{Classification}}

\startdata

1999gs & 0.639 & Ia \\
2000fu & 0.044 & CC \\
2000fv & 0.154 & CC \\
2003bq & 0.066 & CC \\
2003cm & 0.067 & CC \\
2004bt & 0.102 & CC \\
2005lv & 0.162 & CC \\
2006A  & 0.497 & CC \\
2006dz & 0.611 & Ia \\
2008hl & 0.213 & CC

\enddata

\end{deluxetable}

Of the 10 unclassified SNe, 8 are classified as core-collapse
(although SN~2006A, with $p = 0.497$, is effectively undetermined).
Interestingly, SN~2006dz, which {\it galsnid} classifies as a SN~Ia,
was originally identified in template-subtracted images of another SN,
SN~2006br \citep{Contreras06}.  The SN was identified after maximum
brightness, and the SN appears to be heavily dust reddened.
Nonetheless, the light curves are consistent with being a SN~Ia.

Summing the {\it galsnid} $p$ values for the SNe classified as
core-collapse and $1-p$ for those classified as SNe~Ia, we can
estimate the number of incorrectly classified SNe in this sample.  For
the spectroscopically unclassified LOSS sample, we expect incorrect
classifications for 2.055 SNe out of 10 SNe.

\subsection{Further Improvements to Galsnid}\label{ss:improve}

The current {\it galsnid} method is presented mainly as a proof of
concept.  There are several improvements one should make before using
{\it galsnid} for specific robust scientific results.

The current methodology of {\it galsnid} presumes that all host-galaxy
properties are uncorrelated.  This is clearly incorrect.  As a result,
we have some information about other parameters with a single
measurement.  Specifically, color, luminosity, and morphology are
correlated.  Taking these correlations into account should improve our
inference.

There are a number of additional parameters that one could measure for
a host galaxy.  For the LOSS sample, we simply use morphology, a
single color, a single luminosity, an effective offset, and a pixel
ranking.  Adding additional photometry in several bands should improve
classification.  Deriving physical quantities such as star-formation
rates and masses from such data may also provide more robust
classifications.

Adding data from spectroscopy should also improve classification.
Specifically, emission line luminosity \citep[to measure
star-formation rates; e.g.,][]{Meyers12}, line diagnostics (to
determine possible AGN contribution to photometry), velocity
dispersion (to measure a mass), and metallicity could all be important
discriminants.  Additional data such as \ion{H}{1} measurements could
be useful, but perhaps difficult to obtain for large samples.

One could also possibly include other environmental information in a
classifier.  The density of the galactic environment may affect the
relative rates of SNe.  Perhaps close companions are a good indication
of a recent interaction which triggered a burst of star formation.

Future investigations should attempt to provide a broader set of
observations from which classifications can be made.

\subsection{Combining Classifications}

Using host-galaxy properties to classify SNe has several distinct
advantages over light-curve analyses.  It does not use any light-curve
information, so samples will not be biased based on expectation of
light-curve behavior.  Almost all host-galaxy data can be obtained
after the SN has faded.  In particular, high-resolution imaging,
spectroscopy, or additional photometry can be obtained post facto.

Since {\it galsnid} is independent from any light-curve classifier,
one can use SN~Ia samples defined by both techniques to examine
systematics introduced by either method.

Host-galaxy data can also be combined with photometry-only
classification, and one can implement hybrid approaches.  Since {\it
  galsnid} produces a probability density function for each SN, it
would be trivial to naively combine the output of {\it galsnid} with
any other similar output.  However, some SN properties are correlated
with host-galaxy properties.  For instance, SNe~Ibc tend to come from
brighter galactic positions than SNe~II and SNe~Ia hosted in
ellipticals tend to be lower luminosity than those hosted in spirals.
Therefore, a more careful approach to combining different methods
should be used.

In addition to classifications made purely on host-galaxy or
light-curve properties, one could use hybrid measurements.  For
instance the peak luminosity of a SN compared to the luminosity of its
host galaxy or the relative colors of a SN and its host galaxy could
be useful indicators.

\subsection{Redshift Evolution}

The relative fraction of SN classes changes with redshift, with the
SN~Ia fraction decreasing with redshift to at least $z \approx 1$.
Similarly, galaxy properties change, on average, over redshift ranges
of interest.  It is not known if the fractions over a small parameter
range change.  For instance, it is reasonable to assume that the SN~Ia
fraction in ellipticals has relatively little evolution.  The
fractions in other small bins might also stay the same while the
underlying galaxy population is changing with redshift.

The assumption that there is little evolution in the relative
fractions for small parameter ranges should be tested with data.
However, even with this assumption, one needs to account for the
overall evolution of the galaxy population for high-redshift samples.
A simple approach is to have a prior for the overall (observed)
fraction as a function of redshift.  Such a prior can both be
determined observationally and through simulations.

However, if one uses a threshold of a particular {\it galsnid}
probability to separate classes, the effect of the prior is minimized.
That is, the classification of a particular SN is only affected if the
different prior would cause the {\it galsnid} probability of that
object to cross the threshold.  For example, if there are objects with
$p = 0.99$, 0.95, 0.9, 0.8, and 0.6 with $P({\rm Ia}) = 0.5$, changing
the prior to $P({\rm Ia}) = 0.4$ will result in probabilities of $p =
0.985$, 0.93, 0.86, 0.73, and 0.5, respectively.  If the threshold
were $p = 0.94$, only one of the example SNe would have had their
classification changed with the different priors.  This example also
demonstrates that objects with $p \approx 0.5$ are more affected by
the prior than those close to zero or one.

\subsection{When Not to Use Galsnid}

The {\it galsnid} method can produce relatively clean samples of
particular SN classes.  However, these samples can be highly biased
subsamples of the underlying SN class.  For instance, when choosing a
SN~Ia sample, SNe with elliptical hosts will be much more likely to be
included than those in spirals.  But since SN~Ia properties such as
luminosity are correlated with host-galaxy properties
\citep[e.g.,][]{Hicken09:lc}, a {\it galsnid}-defined sample will
likely be biased to lower luminosity SNe.  Similar biases are also
introduced by light-curve classifiers (SNe~Ia with light curves more
like the templates and less like core-collapse SNe are more likely to
be included); however, the {\it galsnid} biases may be harder to
properly model.  Similarly, as seen in Section~\ref{ss:sdss},
cosmological analyses {\it could} be biased if correlations between
host-galaxy properties and Hubble residuals are not removed.  Again,
these biases also apply to light-curve classifiers which are more
likely to select SNe with particular light-curve properties as SNe~Ia
and if those properties (e.g., color) correlate with Hubble residuals
\citep{Scolnic13}.

As a result, one must be careful in choosing appropriate applications
for {\it galsnid} samples.  Clearly, investigations of host properties
of a given class should not be performed on a {\it galsnid} sample.
If one were to use a {\it galsnid} sample to determine SN rates as a
function of redshift, careful attention to the prior is required.
Additionally, SNe with particularly large offsets could have
misidentified host galaxies.  Although this should not affect many
SNe, {\it galsnid} may not provide representative samples specifically
designed to identify such objects.


\section{Conclusions}\label{s:conc}

We have introduced a method for classifying SNe using only galaxy
data.  This method relies on the fact that different SN classes come
from different stellar populations.  Using the LOSS sample, we
estimate the probabilities that particular SN classes have specific
host-galaxy properties and the probabilities that galaxies with
particular properties have particular SN classes.  We define an
algorithm, {\it galsnid}, that combines the host-galaxy data to
determine the Bayesian posterior probability that a given SN is of a
particular class.

We have tested {\it galsnid} in a variety of ways, and have determined
that it provides robust, reliable classifications under many different
scenarios.  We find that of the quantities examined here, morphology
had the most discriminating power.  We have shown that {\it galsnid}
is effective at building relatively pure samples of particular SN
classes, and can be helpful for building samples for SN~Ia cosmology.
We also demonstrated some additional applications for {\it galsnid},
including separating various subclasses.

Past (SDSS), current (Pan-STARRS; PTF) and future SN surveys (Dark
Energy Survey; the Large Synoptic Survey Telescope) should have deep
imaging of all SN host galaxies as a result of the nominal survey.
These data could be used for classification with {\it galsnid} without
additional observations.  However, a relatively small spectroscopic
campaign could provide detailed information that should improve
classifications beyond those presented here.  Moreover,
high-resolution adaptive-optics or {\it Hubble Space Telescope}
imaging could significantly improve any classification by allowing a
precise morphological classification.

Additional improvements to {\it galsnid} could be achieved by taking
into account additional galaxy data, properly handling correlated
data, joining galaxy and SN data, and combining {\it galsnid} results
with those of photometry-based SN classifiers.

\begin{acknowledgments} 

\bigskip

We thank D.\ Scolnic, R.\ Kessler, M.\ Sako, K.\ Barbary, and the
anonymous referee for useful discussions and comments.

Supernova research at Harvard is supported in part by NSF grant
AST-1211196.

Funding for SDSS-III has been provided by the Alfred P.\ Sloan
Foundation, the Participating Institutions, the National Science
Foundation, and the U.S.\ Department of Energy Office of Science. The
SDSS-III web site is http://www.sdss3.org/.

SDSS-III is managed by the Astrophysical Research Consortium for the
Participating Institutions of the SDSS-III Collaboration including the
University of Arizona, the Brazilian Participation Group, Brookhaven
National Laboratory, Carnegie Mellon University, University of
Florida, the French Participation Group, the German Participation
Group, Harvard University, the Instituto de Astrofisica de Canarias,
the Michigan State/Notre Dame/JINA Participation Group, Johns Hopkins
University, Lawrence Berkeley National Laboratory, Max Planck
Institute for Astrophysics, Max Planck Institute for Extraterrestrial
Physics, New Mexico State University, New York University, Ohio State
University, Pennsylvania State University, University of Portsmouth,
Princeton University, the Spanish Participation Group, University of
Tokyo, University of Utah, Vanderbilt University, University of
Virginia, University of Washington, and Yale University.

This paper uses data accessed through the Weizmann Interactive
Supernova data REPository (WISeREP) --
www.weizmann.ac.il/astrophysics/wiserep .

\end{acknowledgments}

\bibliographystyle{fapj}
\bibliography{../astro_refs}


\end{document}